\documentclass[twocolumn,floatfix,preprintnumbers,amsmath,amssymb,nofootinbib]{revtex4}
\usepackage{graphicx}
\usepackage{amsmath}
\usepackage{psfrag}

\newcommand{\fex}{{\it e.g.}}

\newcommand{\cf}{{\it cf.}}
\newcommand{\keV}{\,\text{keV}}
\newcommand{\GeV}{\,\text{GeV}}
\newcommand{\TeV}{\,\text{TeV}}
\newcommand{\cm}{\,\text{cm}}
\newcommand{\gcmcm}{\,\text{g}\,\text{cm}^{-2}}
\newcommand{\s}{\,\text{s}}
\newcommand{\str}{\,\text{str}}
\newcommand{\kpc}{\,\text{kpc}}
\newcommand{\Mpc}{\,\text{Mpc}}
\newcommand{\Gpc}{\,\text{Gpc}}

\begin{document}

\preprint{DESY 09-134, TUM-HEP 736/09}

\title{Detecting Gamma-Ray Anisotropies from Decaying Dark Matter:\\ 
Prospects for Fermi LAT}

\author{Alejandro Ibarra}
\email{alejandro.ibarra@ph.tum.de}
\author{David Tran}
\email{david.tran@ph.tum.de}
\affiliation{
Physik-Department T30d, Technische Universit\"at M\"unchen,\\
James-Franck-Stra\ss e, 85748 Garching, Germany.}
\author{Christoph Weniger}
\email{christoph.weniger@desy.de}
\affiliation{
Deutsches Elektronen-Synchrotron DESY,\\
Notkestra\ss e 85, 22607 Hamburg, Germany.}

\begin{abstract}
  Decaying dark matter particles could be indirectly detected as an excess
  over a simple power law in the energy spectrum of the diffuse extragalactic
  gamma-ray background. Furthermore, since the Earth is not located at the
  center of the Galactic dark matter halo, the exotic contribution from dark
  matter decay to the diffuse gamma-ray flux is expected to be anisotropic,
  offering a complementary method for the indirect search for decaying dark
  matter particles. In this paper we discuss in detail the expected
  dipole-like anisotropies in the dark matter signal, taking also into account
  the radiation from inverse Compton scattering of electrons and positrons
  from dark matter decay. A different source for anisotropies in the gamma-ray
  flux are the dark matter density fluctuations on cosmic scales. We calculate
  the corresponding angular power spectrum of the gamma-ray flux and comment
  on observational prospects. Finally, we calculate the expected anisotropies
  for the decaying dark matter scenarios that can reproduce the
  electron/positron excesses reported by PAMELA and the Fermi LAT, and we
  estimate the prospects for detecting the predicted gamma-ray anisotropy in
  the near future. 
\end{abstract}

\maketitle

\section{Introduction}
The full-sky observations of gamma rays undertaken in the 1990s by the
Energetic Gamma Ray Experiment Telescope (EGRET) revealed a map which is
highly anisotropic. It showed a number of resolved
sources~\cite{Nolan:1995aj}, such as blazars, as well as an unresolved
component, which can be attributed almost entirely to Galactic emission. The
mechanisms which produce the diffuse Galactic emission (inverse Compton
scattering, $\pi^0$ production and bremsstrahlung) are well understood.
However, the actual intensity of each of these contributions depends on many
details of the propagation of cosmic rays in the Galaxy, including the
interstellar radiation and magnetic fields and the Galactic gas distribution.
On the other hand, over the last years a picture of cosmic-ray propagation has
emerged which can account for the observed abundances of {\it almost} all
secondary cosmic-ray species. Remarkably, this same propagation model can
reproduce, using the interstellar gas and radiation field distributions and a
Galactic magnetic field inferred from observations, the full-sky gamma-ray map
with rather good accuracy. However, this requires the introduction of an
additional, a priori undetermined, diffuse component which is postulated to be
of extragalactic origin and thus isotropic. Its energy spectrum is determined
by observations at high Galactic latitudes. 

Recently it has become apparent that the state-of-the-art propagation models
fail to reproduce the measurements of the \textit{positron fraction} at
energies larger than 10 GeV. Namely, the secondary positron flux calculated
with current propagation models, together with the total
electron-plus-positron spectrum measured by the Fermi LAT, yields a positron
fraction which monotonically decreases with energy, whereas the PAMELA
observations reveal that above 10 GeV the positron fraction increases with
energy~\cite{A08, Abdo:2009zk, Grasso:2009ma}. This puzzling behavior, already
suggested in the past by a series of experiments such as HEAT~\cite{B97}, has
been interpreted as evidence for a primary component of the positron flux
(possibly accompanied by an identical electron flux). Furthermore, the
conventional propagation model adopted by the Fermi collaboration, dubbed
``model 0'', which fits well the low-energy data points of the total
electron-plus-positron flux and the positron fraction, fails to reproduce the
observations at higher energies, reinforcing the necessity of an exotic source
of primary electrons and positrons.

The most common astrophysical explanation of the electron/positron excesses is
the electron-positron pair production by the interactions of high-energy
photons in the strong magnetic field of nearby pulsars, such as Geminga or
Monogem~\cite{Grimani:2004qm, Profumo08, HBS09, Malyshev:2009tw}.  However,
this interpretation requires a rather large percentage of the total spin-down
power injected in the form of electron-positron pairs, namely about $40\%$,
and a large cutoff of the electron/positron energy spectrum at about 1 TeV.
Alternatively, the electron/positron excesses could be explained by the
combined emission of both nearby and distant pulsars, this solution requiring
a percentage of spin-down power ranging between 10 $-$ 30\% and again a large
cutoff in the energy spectrum, 800 $-$ 1400 GeV~\cite{Grasso:2009ma}.

An arguably more exciting explanation of the cosmic-ray electron/positron
excesses is the possibility that the electrons and positrons are produced in
the annihilation~\cite{Bergstrom:2009fa, Meade:2009iu, Bi:2009am, Chen:2009ab}
or the decay~\cite{Ibarra:2009dr, Ibarra:2009bm, IT08a, CTY08, Kohri:2009yn}
of dark matter particles. Despite the overwhelming gravitational evidence for
the existence of non-baryonic dark matter, which makes up more than one-fifth
of the energy budget of the Universe~\cite{D09}, very little is known about
its particle nature (see Ref.~\cite{BHS05} for a review). The most popular
type of dark matter candidate, the weakly interacting massive particle (WIMP),
can naturally reproduce the observed dark matter abundance due to effective
self-annihilation in the early Universe, after being in thermal equilibrium
with the baryons before. Today, this same annihilation process is expected to
produce a possibly observable contribution to the measured cosmic-ray fluxes
on Earth. Detection of such an indirect signal would be the first
non-gravitational evidence for dark matter, with paramount importance to the
understanding of its nature.

A lot of effort has been made to study the prospects and predictions of
cosmic-ray signatures from annihilating dark matter, see \fex~\cite{UBE+02,
PLB+09, RU09}. However, this is not the only possibility for the indirect
detection of dark matter. Many dark matter models predict that the dark matter
particle is unstable and decays with very long lifetimes~\cite{BCH+07, IRW08,
Ibarra:2009bm, CTY08, Kohri:2009yn}. If the decays occur at a sufficiently
large rate, the decay products could be observable as an exotic contribution
to the high energy cosmic ray fluxes of gamma rays, electrons, positrons,
antiprotons, neutrinos or antideuterons~\cite{Buckley:2009kw, Ibarra:2009tn,
Kadastik:2009ts, Spolyar:2009kx, IMM08b, IT08b, Ruderman:2009ta}. Among these,
the gamma-ray channel is probably the most important to study, due to its
sensitivity to far-distant sources and its potential to discriminate between
signals from annihilating or decaying dark matter and astrophysical sources.

In this paper we will present a detailed study of the peculiar predictions for
gamma rays from decaying dark matter. We will concentrate on their angular
anisotropies on large and small scales. These can be used to discriminate this
component from other contributions to the extragalactic diffuse emission. It
turns out that the predictions for decaying dark matter are much more robust
than the ones for annihilation, which makes this scenario very predictive and
easier to confirm or falsify~\cite{BBC+07}. Secondly, we will discuss the
prospects to see such signals in the upcoming Fermi LAT gamma-ray data, and we
will apply our results to the decaying dark matter explanation of the
electron/positron excess.

This paper is organized as follows: In section~\ref{sec:Gammas} we will review
the basic concepts about production, propagation and absorption of gamma rays
from dark matter decay. In section~\ref{sec:Fermi} we will calculate the
gamma-ray anisotropy expected from the decay of dark matter particles on large
angular scales. We will show in particular the expected anisotropy in
scenarios which can explain the electron/positron excesses observed by PAMELA
and the Fermi LAT, and we will argue that this anisotropy should be seen by
the Fermi LAT. In section~\ref{sec:power} we will calculate the angular power
spectrum of gamma rays from decaying dark matter on small angular scales and
comment on its observational prospects. Lastly, in
section~\ref{sec:Conclusion}, we will present our conclusions. We also present
two appendices in which we discuss statistical properties of the large-scale
anisotropy and the general observational strategy for gamma rays from dark
matter decay.

\section{Gamma rays from dark matter decay}
\label{sec:Gammas}
The total gamma-ray flux from dark matter decay receives several
contributions. The first one stems from the decay of dark matter particles in
the Milky Way halo and reads
\begin{equation}
  \frac{dJ_\text{halo}}{dE_\gamma}(l,b) = 
  \frac{1}{4\pi\,m_\text{dm}\,\tau_\text{dm}} 
  \frac{dN_\gamma}{dE_\gamma}
  \int_0^\infty ds\; 
  \rho_{\rm halo}[r(s,l,b)] \;,
  \label{halo-flux}
\end{equation}
where $dN_\gamma/dE_\gamma$ is the energy spectrum of gamma rays produced in
the decay of a dark matter particle and $\rho_{\rm halo}(r)$ is the density
profile of dark matter particles in our Galaxy, as a function of the distance
from the Galactic center, $r$. The received gamma-ray flux depends on the
Galactic coordinates, longitude $l$ and latitude $b$, and is given by a
line-of-sight integral over the parameter $s$, which is related to $r$ by
\begin{equation}\label{galcoord}
  r(s,l,b) = \sqrt{s^2+R^2_\odot-2 s R_\odot \cos b\cos l}\;.
\end{equation}
Here, $R_\odot=8.3\kpc$ denotes the distance of the Sun to the Galactic center
\cite{G09a}.

The latest $N$-body simulations favor the Einasto density profile
\cite{Navarro:2003ew,Springel:2008cc,PLB+09}
\begin{equation}
  \rho^\text{Einasto}_\text{halo}(r) \propto
  \exp\left[-\frac{2}{\alpha}\left(\left(\frac{r}{r_s}\right)^\alpha - 1
  \right)\right]\;,
\end{equation}
which we use throughout the work when not stated otherwise, and for which we
adopt $\alpha = 0.17$ and the scale radius $r_s = 20\kpc$. For comparison we
will also show results for the much shallower isothermal profile
\begin{equation}
  \rho^\text{isothermal}_\text{halo}(r)\propto
  \frac{1}{r^2+r_s^2}
\end{equation}
with $r_s = 3.5\kpc$. We use the local dark matter density as determined in
Ref.~\cite{CU09}, $\rho_\odot=0.385\GeV\cm^{-3}$, to normalize the profiles at
the position of the Sun. Related uncertainties and their impact on our results
will be discussed below.\\

The electrons and positrons that can be produced in the decay of dark matter
particles, and which may be the origin of the PAMELA and Fermi LAT anomalies,
also generate a contribution to the total gamma-ray flux through their inverse
Compton scattering on the interstellar radiation field (ISRF), which includes
the CMB, thermal dust radiation and starlight. Recently, ICS radiation in
connection with the PAMELA excess was discussed in Refs.~\cite{Cirelli:2009vg,
Ishiwata:2009dk, Kistler:2009xf, Profumo:2009uf, Zhang:2008tb,
Borriello:2009fa}; a pedagogical review can be found in
Ref.~\cite{Blumenthal:1970gc}. Furthermore, the interactions of energetic
electrons and positrons with the Galactic magnetic field produce synchrotron
radiation in the radio band with frequencies ${\cal O}(0.1 - 100~{\rm GHz})$,
which could also be observed (see \fex~Ref.~\cite{Zhang:2009pr,
Ishiwata:2008qy}).

The production rate of gamma rays with energy $E_\gamma$ at the position
${\vec r}$ of the Galaxy, due to inverse Compton scattering of dark matter
electrons (or positrons) with number density $f_{e^\pm}(E_e,{\vec r})$ on
photons of the ISRF with number density $f_{\rm ISRF}(\epsilon,\vec{r})$, is
given by
\begin{equation}
  \begin{split}
    &\frac{dR^{\rm IC}_\gamma(\vec r)}{d E_\gamma}=\\
    &\int_0^\infty d\epsilon  
    \int_{m_e}^{\infty} dE_e\; \frac{d\sigma^{\rm
    IC}(E_e,\epsilon)}{dE_\gamma} f_{e^\pm}(E_e,{\vec r})
    f_{\rm ISRF}(\epsilon,\vec{r})\;.
  \end{split}
  \label{eqn:IC-rate}
\end{equation}
Here, $d\sigma^\text{IC}/dE_\gamma$ denotes the differential cross section of
inverse Compton scattering of an electron with energy $E_e$, where an ISRF
photon with energy $\epsilon$ is up-scattered to energies between $E_\gamma$
and $E_\gamma+dE_\gamma$. It can be derived from the Klein-Nishina formula and
is given by
\begin{equation}
  \begin{split}
    &\frac{d\sigma^{\rm IC}(E_e,\epsilon)}{dE_\gamma}=
    \frac{3}{4}\frac{\sigma_{\rm T}}{\gamma_e^2\, \epsilon}\,\times \\
    &\times\left[2q\ln q + 1 + q - 2q^2
      +\frac{1}{2}\frac{(q\Gamma)^2}{1+q\Gamma}(1-q)\right],
  \end{split}
  \label{eqn:ICrate}
\end{equation}
where $\sigma_{\rm T}=0.67\,{\rm barn}$ denotes the Compton scattering cross
section in the Thomson limit, $\gamma_e\equiv E_e/m_e$ is the Lorentz factor
of the electron, $m_e=511\keV$ is the electron mass, and we have defined
$\Gamma\equiv 4\gamma_e\epsilon/m_e$ and $q\equiv
E_\gamma/\Gamma(E_e-E_\gamma)$. Eq.~\eqref{eqn:ICrate} holds in the limit
where $\epsilon,m_e\ll E_e$, and kinematics and the neglect of down-scattering
require that $\epsilon\leq E_\gamma \leq (1/E_e + 1/4\gamma_e^2 \epsilon)^{-1}
\equiv E_\gamma^\text{max}$. In the calculations we will further assume that 
the photon and electron fields are isotropic; taking into account the 
anisotropy of the photons, which are mainly produced in the Galactic disk, 
would give $\mathcal{O}(10\%-20\%)$ corrections to the ICS 
fluxes~\cite{Moskalenko:1998gw}.

The gamma-ray flux from ICS that is received at Earth reads
\begin{equation}
  \frac{dJ_\text{halo-IC}}{dE_\gamma}(l,b) =
  2\cdot\frac{1}{4\pi} 
  \int_0^\infty ds\;\frac{dR^{\rm IC}_\gamma[r(s,l,b)]}{d
  E_\gamma}\;,
  \label{eqn:IC-flux-halo}
\end{equation}
where the factor of 2 takes into account the fact that both dark matter
electrons and positrons contribute equally to the total flux of gamma rays.

For the number density of ISRF photons we will use results from
Ref.~\cite{Porter:2005qx}. On the other hand, the number density of electrons
and positrons from dark matter decay follows from solving the appropriate
transport equation, which incorporates the effects of diffusion,
reacceleration and convection in the Galactic magnetic field and energy losses
due to synchrotron emission and inverse Compton scattering on the ISRF (see
\fex~\cite{MS98, SMP07,DLD+08}). However, at higher energies above a few
$10\GeV$ the transport equation is dominated by the energy loss terms, and the
number density of electrons and positrons can be approximated by
\begin{equation}
  \begin{split}
    f_{e^\pm}(E_e,\vec{r})&=\frac{1}{b(E_e,\vec{r})}
    \frac{\rho_\text{halo}(\vec{r})}{m_\text{dm}\,\tau_\text{dm}}
    \int_{E_e}^\infty d\tilde{E_e}\; 
    \frac{dN_{e^\pm}}{d\tilde{E_e}}\;.
  \end{split}
  \label{eqn:IC-electrons}
\end{equation}
Here, $b(E_\text{e},\vec{r})$ accounts for the energy losses and contains a
part that comes from ICS on the ISRF, and a part that comes from synchrotron
losses in the Galactic magnetic field, $b=b_\text{ICS}+b_\text{syn}$. We set
$f_{e^\pm}=0$ outside of the diffusion zone, which we model by a cylinder of
half-height $L=3\kpc$ and radius $R=20\kpc$.\footnote{For some sample decay channels
we have cross-checked with \textsc{Galprop} v50p, using appropriately modified
versions of the model \textsc{50p\_599278} (which adopts a diffusive halo with
$L=4\kpc$) and of the annihilation package, that our approximations give
correct ICS gamma-ray sky maps at the $30\%$ level everywhere in the sky for
gamma-ray energies above $1\GeV$.} The impact of a variation of the height of
the diffusive halo on our results will be discussed below.

The part of the energy loss that is due to ICS is given by
\begin{equation}
  \begin{split}
    &b_\text{ICS}(E_e,\vec{r})=\\&\int_0^\infty d\epsilon
    \int_{\epsilon}^{E_\gamma^\text{max}} dE_\gamma\,
    (E_\gamma-\epsilon)\, \frac{d\sigma^{\rm
    IC}(E_e,\epsilon)}{dE_\gamma}f_{\rm ISRF}(\epsilon,\vec{r})\;.
  \end{split}
\end{equation}
For electron energies $E_e=1\GeV$, $b_\text{ICS}$ ranges between
$4.1\times10^{-17}\GeV\s^{-1}$ and $1.9\times10^{-15}\GeV\s^{-1}$, depending
on $\vec{r}$. At higher energies $b_\text{ICS}$ approximately scales like
$\sim E_e^2$. On the other hand, the synchrotron loss part reads
\begin{equation}
  b_\text{syn}(E_e,\vec{r})=\frac{4}{3}\sigma_\text{T}
  \gamma_e^2 \frac{B^2}{2}\;,
\end{equation}
where $B^2/2$ is the energy density of the Galactic magnetic field, and we set
$B=6\,\mu\text{G}\exp(-|z|/5\kpc-r/20\kpc)$ for
definiteness~\cite{Zhang:2009pr}. At position of the Sun this yields an energy
loss of $b_\text{syn}\simeq 4.0\times10^{-17}(E_e/\GeV)^2\GeV\s^{-1}$.

Substituting Eq.~\eqref{eqn:IC-electrons} into Eq.~\eqref{eqn:IC-rate}, the
flux from inverse Compton scattering of electrons and positrons from 
dark matter on the Galactic radiation field can be calculated.\\

In addition to the gamma-ray fluxes that originate from the decay of dark
matter particles in the Milky Way halo, there exists a largely isotropic
contribution generated by the decay of dark matter particles at cosmological
distances.  Analogously to the Milky Way component, the latter receives
contributions from the direct decay of dark matter particles into photons, and
from the gamma rays produced by the inverse Compton scattering of dark matter
electrons and positrons on the intergalactic radiation field.

The direct decay of dark matter particles at cosmological distances produces a
gamma-ray flux that is given by
\begin{equation}
  \begin{split}
    &{}\frac{dJ_\text{eg}}{dE_\gamma} =
    \frac{\Omega_\text{dm}\rho_\text{c}}{4\pi m_\text{dm} \tau_\text{dm}}
    \times \\ &\times\int_0^\infty dz
    \frac{1}{H(z)}
    \;\frac{dN_\gamma}{dE_\gamma}\left[(z+1)E_\gamma\right]
    \;e^{-\tau(E_\gamma,z)}\;,
  \end{split}
  \label{eqn:GammaFluxEG}
\end{equation}
where $H(z)=H_0 \sqrt{\Omega_\Lambda+\Omega_\text{m}(z+1)^3}$ is the Hubble
expansion rate as a function of redshift $z$, and
$\rho_\text{c}=5.5\times10^{-6}\GeV/\cm^3$ denotes the critical density of the
Universe. Throughout this work we assume a $\Lambda$CDM cosmology with
parameters $\Omega_\Lambda=0.74$, $\Omega_\text{m}=0.26$,
$\Omega_\text{dm}=0.21$ and $h\equiv
H_0/100\,\text{km}\;\text{s}^{-1}\,\text{Mpc}^{-1}=0.72$, as derived from the
five-year WMAP data~\cite{D09}.

The extragalactic ICS radiation of electrons and positrons from dark matter is
expected to come mainly from scattering on the CMB~\cite{Ishiwata:2009dk}.
Note that there is a similar component that comes from electrons and positrons
produced in the Milky Way halo, but outside of the diffusion zone (see
Ref.~\cite{Kistler:2009xf} for the analogous case of annihilation). For dark
matter particles with masses below $3-5\TeV$ these components are all expected
to become relevant only for gamma-ray energies $E_\gamma\lesssim10\GeV$. We
will neglected extragalactic ICS radiation from dark matter decay in the
present work. Note, however, that this radiation would be largely isotropic
and would somewhat reduce the overall anisotropy of the ICS radiation at these
lower energies.\\

\begin{figure}[h!]
  \begin{center}
    \psfrag{energy}[]{\scriptsize $E_\gamma [\text{GeV}]$}
    \psfrag{redshift}[]{\scriptsize $z$}
    \psfrag{tau10}{\scriptsize $\tau=10$}
    \psfrag{tau1}{\scriptsize $\tau=1$}
    \psfrag{tau01}[]{\scriptsize $\tau=0.1$}
    \psfrag{FastEvolution}[lc]{\scriptsize fast evolution}
    \psfrag{Baseline}[lc]{\scriptsize baseline}
    \includegraphics{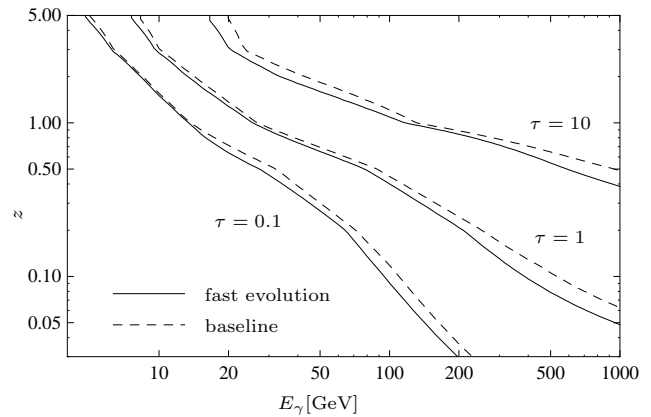}
  \end{center}
  \caption{Isocontours of the optical depth $\tau(E_\gamma,z)$ of gamma-ray
  photons, emitted at redshift $z$ and observed at Earth with energy
  $E_\gamma$. We show results for two different models of the intergalactic
  background light~\cite{SMS06a}. Throughout this work, we will adopt the
  ``fast evolution'' model.}
  \label{fig:OD}
\end{figure}

In Eq.~\eqref{eqn:GammaFluxEG} we included an attenuation factor for the
gamma-ray flux, which incorporates the effects of electron-positron pair
production by collisions of gamma rays from dark matter decay with the
extragalactic background light emitted by galaxies in the ultraviolet, optical
and infrared frequencies~\cite{Gould:1967zzb}. The attenuation factor is
determined by the optical depth $\tau(E_\gamma,z)$, for which we will use the
results from~\cite{SMS06a} throughout this work.\footnote{In
Ref.~\cite{SMS06a} the optical depth is calculated for redshifts $z<5$.
Following Ref.~\cite{CBH+08}, we assume that the optical depth does not
increase beyond $z=5$ and set $\tau(E_\gamma)_{z>5}=\tau(E_\gamma)_{z=5}$.} In
Fig.~\ref{fig:OD} we show isocontours of the optical depth in the redshift
vs.~energy plane. It is apparent from the plot that gamma rays with high
energies around $1\TeV$ are strongly attenuated and come mainly from redshifts
$z\lesssim 0.05$. On the other hand, the flux of gamma rays originating from
the decay of dark matter particles in the Galactic halo is barely attenuated
by pair production on the ISRF at energies below 10
TeV~\cite{Moskalenko:2005ng}.

\begin{figure}[h!]
  \begin{center}
    \psfrag{flux}[]{\scriptsize $\propto J\,
    [ph\;\text{cm}^{-2}\text{s}^{-1}\text{str}^{-1}]$}
    \psfrag{angle}[]{\scriptsize $\psi\,[\text{degree}]$}
    \psfrag{halo}[]{\scriptsize halo2}
    \includegraphics{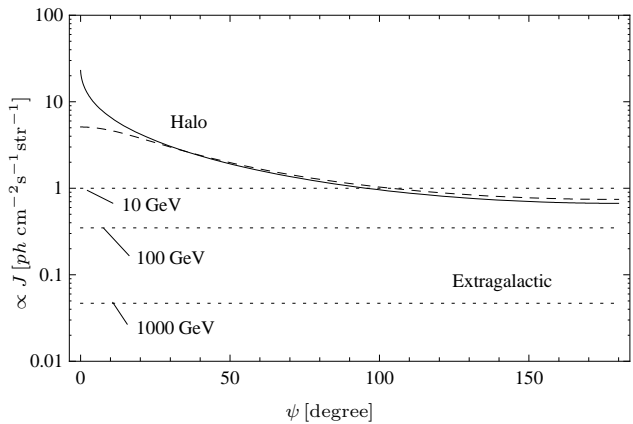}
  \end{center}
  \caption{Angular profile of the gamma-ray signal from dark matter decay as
  function of the angle $\psi$ to the center of the galaxy. The \textit{solid}
  (\textit{dashed}) line shows the contribution from decay in the Milky Way
  halo, assuming the Einasto (isothermal) profile. Extragalactic contributions
  are shown in \textit{dotted} lines for the case that dark matter decay
  produces a monoenergetic line with energies between $E'_\gamma = 10$ and
  $1000\GeV$. The fluxes are integrated over energy and normalized to the size
  of the extragalactic component when absorption is neglected.}
  \label{fig:AngDist}
\end{figure}

To compare the sizes of the halo and the extragalactic component of gamma rays
from decaying dark matter (neglecting ICS radiation for simplicity), we show
in Fig.~\ref{fig:AngDist} the total flux of photons, integrated over all
energies, as a function of the angular distance $\psi$ from the Galactic
center, for a dark matter particle which decays producing a monoenergetic
photon with an energy in the range $E'_\gamma\simeq10\GeV-1\TeV$. As apparent
from the figure, the cosmological contributions decrease with energy due to
the attenuation described above, while the radiation profile from decaying
particles in the halo is independent of energy. The halo contribution
typically dominates the total flux independently of the halo profile, except
at low energies $E'_\gamma\lesssim10\GeV$ in the direction of the Galactic
anticenter. The differences in the two dark matter profiles become only
relevant near the Galactic center when $\psi\lesssim 10^\circ$, and at the
Galactic center the flux predicted for the Einasto profile is almost one order
of magnitude larger than the corresponding flux from the isothermal profile.

\section{Large-scale anisotropies in gamma rays from decaying dark matter}
\label{sec:Fermi}
The decay of dark matter particles can produce gamma rays that could be
detected as an exotic contribution to the diffuse extragalactic gamma-ray
background (EGBG). The diffuse extragalactic background at high energies is
believed to be dominated by the emission from unresolved active galactic
nuclei and is expected to approximately follow a simple power law, with an
intensity and index that has to be determined by fitting to the
data~\cite{S98, SMR04, Dermer:2007fg}. Thus, if dark matter particles decay at
a sufficiently fast rate, one generically expects to observe a deviation from
a simple power law in the gamma-ray energy spectrum, which could show up in
experiments like Fermi LAT.

A complementary signature of dark matter decay is the observation of
anisotropies in the EGBG. It is well known that the offset between Sun and
Galactic center causes a peculiar angular dependence in the gamma-ray signal
from dark matter decaying~\cite{BBC+07} (or
annihilating~\cite{CalcaneoRoldan:2000yt, HS07}) in the Milky Way halo, even
if the halo profile itself is isotropic. The halo signal is largest in
direction of the Galactic center and smallest in direction of the Galactic
anticenter (\cf~Fig.~\ref{fig:AngDist}). The observation of an anisotropy that
is aligned in this way would be a strong signal for a contribution from dark
matter, and, on the other hand, its non-observation would provide strong
constraints. Gamma rays from the decay of dark matter particles at
cosmological distances are isotropic and tend to reduce the anisotropy of the
overall signal. This attenuation effect is however always small, due to the
relative weakness of the extragalactic component.\\

To analyze the prospects of detecting a gamma-ray anisotropy from dark matter
decay at the Fermi LAT it is convenient to define the quantity
\begin{equation}
  A_{b_0:b_1}=\frac {\bar{J}_\text{GC}-\bar{J}_\text{GAC}}
  {\bar{J}_\text{GC}+\bar{J}_\text{GAC}} \;,
  \label{eqn:Ani}
\end{equation}
where $\bar{J}_\text{GC}$ and $\bar{J}_\text{GAC}$ in general denote the total
diffuse gamma-ray flux (from dark matter and from astrophysical sources)
integrated over $E_\gamma$ in some energy range, and averaged over the
hemisphere in direction of the Galactic center (GC) and anticenter (GAC),
respectively. Sky regions with small, $|b|<b_0$, or large, $|b|>b_1$, Galactic
latitudes are excluded from the average. 

\begin{table}[h!]
  \centering
  \begin{tabular}{c|c|c}
    \hline\hline
    Sky patch  & \multicolumn{2}{c}{Anisotropy $A_{b_0:b_1}$} \\
    $b_0:b_1$ & Einasto & Isothermal \\\hline\hline
    $10^\circ:90^\circ$ & $0.21-0.36$ & $0.20-0.33$\\
    $10^\circ:20^\circ$ & $0.32-0.50$ & $0.29-0.45$\\
    $20^\circ:60^\circ$ & $0.21-0.35$ & $0.20-0.33$\\
    $60^\circ:90^\circ$ & $0.07-0.13$ & $0.07-0.13$\\\hline\hline
  \end{tabular}
  \caption{Anisotropy of the gamma-ray signal from the decay  $\psi\rightarrow
  \gamma \nu$ of a fermionic dark matter particle $\psi$, after subtracting
  astrophysical contributions. The ranges correspond to the anisotropies from
  gamma-ray lines with energies between $10\GeV$ and $1000\GeV$ in different
  regions of the sky, see Eq.~\eqref{eqn:Ani}.}
  \label{tab:Ani}
\end{table}

As an example, we calculate the anisotropy parameter $A$ in different regions
of the sky for the decay $\psi\rightarrow \gamma \nu$ of a fermionic dark
matter particle $\psi$, after subtracting astrophysical sources (the Galactic
foreground and the extragalactic gamma-ray background), in order to compare
the anisotropy expected {\it purely} from dark matter decay to the anisotropy
expected from the Galactic models (in the rest of the paper, however, we will
consider both sources of gamma rays simultaneously, calculating the anisotropy
of the total flux). In this case, the energy spectrum of gamma rays has two
components: a monoenergetic line from the decay of dark matter particles in
the halo and a redshifted line from decays at cosmological distances (note
that in this decay channel there is no contribution from ICS). We show in
Tab.~\ref{tab:Ani} the values of $A$ for dark matter particles with masses
between 20 GeV and 2 TeV, producing monoenergetic photons with energies 10 GeV
and 1 TeV, respectively. It is interesting that the anisotropy parameter can
be as large as 0.5 for large energies and relatively low latitudes. In the
region defined by $b_0=10^\circ$ and $b_1=90^\circ$ (on which we will
concentrate below), the anisotropy of the ``pure'' dark matter signal ranges
between $0.20$ and $0.36$, with only little dependence on the profile of the
dark matter halo. These values have to be compared with the anisotropies of
the Galactic foreground as predicted by \textsc{Galprop} (see below), which
are considerably smaller, and typically $A\lesssim0.10$ in all the regions
that are shown in Tab.~\ref{tab:Ani}, up to energies above $300\GeV$.
Furthermore, the anisotropies measured by EGRET for energies below $10\GeV$
are consistent with the predictions for the Galactic foreground~\cite{CH05}.\\

From the theoretical point of view, the search for anisotropies in the
gamma-ray flux is a cleaner method for the indirect detection of dark matter
than the search for an excess in the spectrum of the EGBG. As mentioned above,
the genuinely extragalactic flux from active galactic nuclei and other
extragalactic sources is very poorly understood. Thus, it is difficult to make
firm predictions for the total gamma-ray flux in scenarios with decaying dark
matter, even when the particle physics model is specified (namely, the dark
matter mass, lifetime and decay modes). Moreover, there are other potentially
important isotropic contributions to the total flux with an intensity that
cannot be predicted theoretically. For instance, interactions of high energy
cosmic rays with debris in the hypothetical Oort cloud could produce a sizable
gamma-ray flux, provided that the column density is larger than
$10^{-3}\gcmcm$ \cite{Moskalenko:2009tv}. Since all these contributions to the
total flux are perfectly isotropic, they cancel out when calculating the
difference of the fluxes between the Galactic center and the Galactic
anticenter regions.\\

To illustrate the large-scale anisotropy $A$ that could be produced by dark
matter decay, we will show predictions for different dark matter decay
channels and masses, including Galactic foreground radiation and the
astrophysical extragalactic background. As Galactic foreground we take
predictions of the conventional \textsc{Galprop}-model as presented in
Ref.~\cite{SMR04a} (model \textsc{44\_500180}), and as astrophysical EGBG we
will use the following parameterization, which agrees with preliminary results
from Fermi LAT~\cite{Ackermann09}
\begin{equation}
  \frac{dJ_\text{EGBG}}{dE_\gamma}=5.8\times 10^{-7} (\text{GeV}\cm^2\s\str)^{-1}
  \left( \frac{E_\gamma}{1\GeV} \right)^{-2.45}\;.
  \label{eqn:EGBG}
\end{equation}

\begin{figure*}[h!]
  \begin{center}
    \psfrag{energy}[][]{\scriptsize $E_\gamma\,[\text{GeV}]$}
    \psfrag{anisotropy}[][]{\scriptsize $A$}
    \psfrag{flux}[bc][tc]{\scriptsize
    $dJ/dE_\gamma\,[\text{GeV}\text{cm}^{-2}\text{s}^{-1}\text{str}^{-1}]$}
    \psfrag{legendLong}[tc][cr]{
    \begin{minipage}[h]{2cm}\scriptsize
      DM$\rightarrow \tau^+\tau^-$
      \begin{align*}
        m_\text{dm}&=600\GeV\\
        \tau_\text{dm}&=3.5\times10^{27}\s
      \end{align*}
    \end{minipage}}
    \psfrag{legendShort}[tc][cr]{\scriptsize DM$\rightarrow \tau^+\tau^-$}
    \psfrag{1090}[t][c]{\scriptsize$10^\circ:90^\circ$}
    \psfrag{1020}[t][c]{\scriptsize$10^\circ:20^\circ$}
    \psfrag{2060}[t][c]{\scriptsize$20^\circ:60^\circ$}
    \psfrag{6090}[t][c]{\scriptsize$60^\circ:90^\circ$}
    \includegraphics[width=0.88\linewidth]{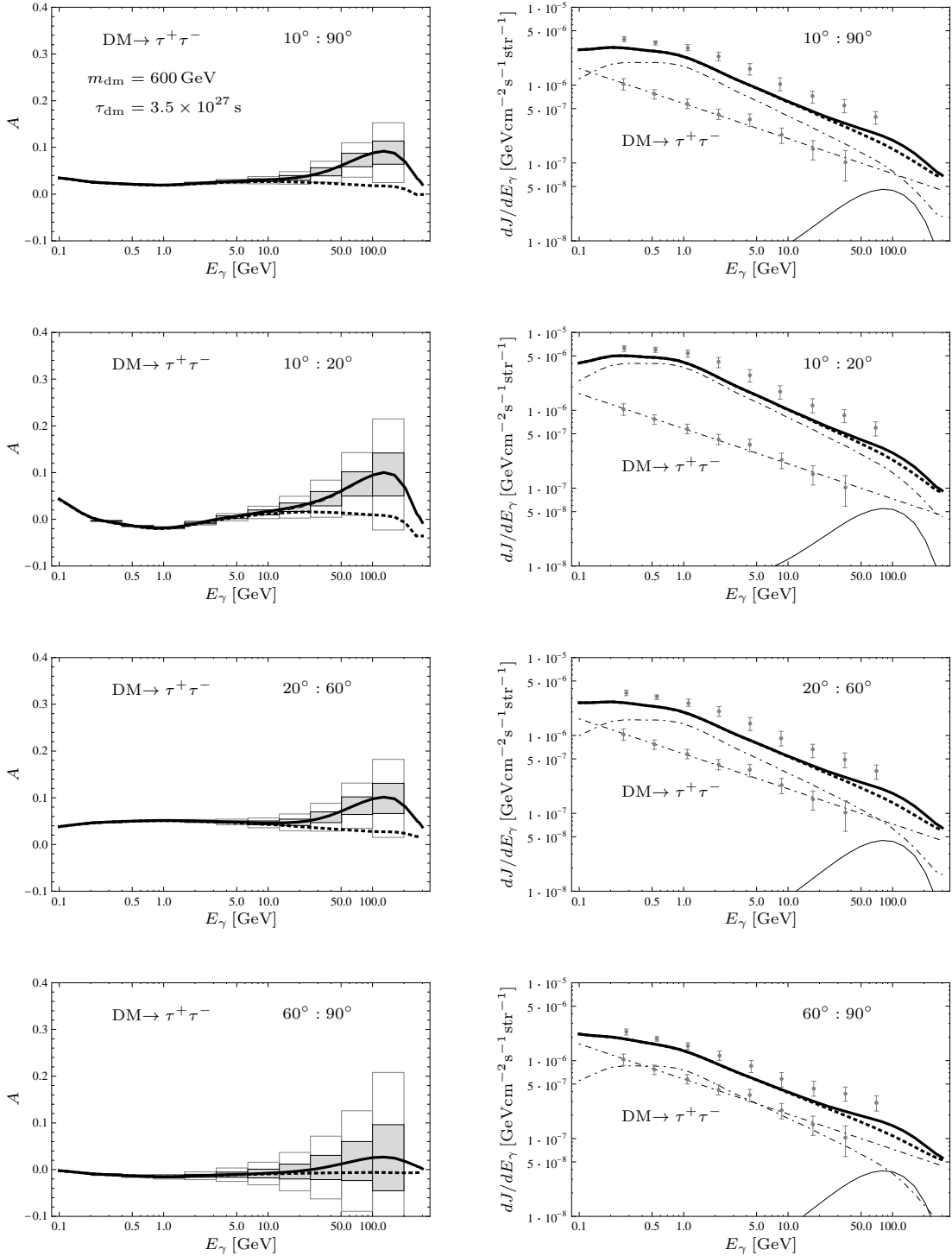}
  \end{center}
  \caption{\textit{Left panels:} Anisotropy of the gamma-ray flux from dark
  matter decay into $\tau^+\tau^-$ pairs as a function of energy. The
  \textit{dotted line} shows the background anisotropy as expected from the
  Galactic foreground, while the \textit{solid line} shows the anisotropy of
  signal + background. We also show the signal + background anisotropy
  neglecting gamma rays from ICS (\textit{dashed}, overlapping with solid
  line). The boxes show estimates of the statistical errors for one-year and
  five-year Fermi LAT observations. \textit{Right panels:} Gamma-ray fluxes
  averaged over all Galactic longitudes as a function of energy. The
  \textit{thin solid line} shows the gamma rays from dark matter decay. The
  two \textit{dash-dotted lines} show the astrophysical EGBG and the Galactic
  foreground separately. The \textit{thick solid line} shows the sum of all
  contributions, whereas the \textit{dotted line} shows the sum without
  contributions from dark matter. From \textit{top to bottom} the different
  panels show predictions for different patches of the sky. The data points
  show preliminary Fermi LAT data~\cite{Ackermann09} for the total diffuse
  flux (\textit{upper points}, in the $10^\circ:90^\circ$ region we averaged
  the presented preliminary results appropriately) and the EGBG (\textit{lower
  points}).}
  \label{fig:tautauAni}
\end{figure*}

We show in the left panels of Fig.~\ref{fig:tautauAni} the predicted
anisotropy of the \textit{total} gamma-ray flux that would be measurable in
different regions of the sky if the dark matter particles decay exclusively
into $\tau^+\tau^-$ pairs. The anisotropy is calculated with taking into
account the galactic foreground and the extragalactic gamma-ray background as
discussed above. The dark matter mass is taken to be $m_\text{dm}=600\GeV$,
and we set the lifetime to $\tau_\text{dm}\simeq 3.5\times10^{27}\s$. The
energy spectra $dN_{\gamma,\text{e}}/dE_{\gamma,\text{e}}$ of the photons,
electrons and positrons produced in the decay are calculated with the event
generator \textsc{Pythia 6.4}~\cite{SMS06}. The lifetime is chosen such that
the gamma-ray fluxes are below and compatible with the EGBG, as demonstrated
in the right panels of Fig.~\ref{fig:tautauAni}, where we also show
preliminary results from the Fermi LAT for comparison.\footnote{Fitting the
preliminary Fermi LAT results with a Galactic foreground model is well beyond
the scope of this paper. Hence, there is a mismatch between the total fluxes
and the data, which does not affect our conclusions.} Furthermore, we note
that the contribution to the local electron and positron fluxes from dark
matter decay is negligible in this scenario. Interestingly, for the adopted
choice of parameters, an anisotropy is predicted that is significantly
different to the one expected from the diffuse Galactic emission in the
conventional \textsc{Galprop} model. Such an anisotropy should, moreover, be
observable by Fermi LAT, as illustrated by the boxes in the figure, which
correspond to our estimates of the one-year and five-year statistical errors
of Fermi LAT, assuming exposures of
$\varepsilon=3\times10^{10}\cm^2\s$~\cite{Ackermann09} and
$\varepsilon=2\times10^{11}\cm^2\s$~\cite{B08}, respectively (for a discussion
about our calculation of the statistical errors see below and
appx.~\ref{sec:dA}). As expected, the size of the anisotropy is largest at low
latitudes $b\lesssim 20^\circ$, and decreases slowly when considering higher
latitudes. On the other hand, the statistical error is smallest in the
sky-patch shown in the upper panels of Fig.~\ref{fig:tautauAni}, where the
whole sky with $|b|\geq10^\circ$ is included. The effects of ICS radiation are
negligible in the present case of decay into $\tau^+\tau^-$ pairs, since the
electrons and positrons produced in the subsequent decay of the taus have only
relatively small energies.

\begin{figure*}[h!]
  \begin{center}
    \psfrag{energy}[][]{\scriptsize $E_\gamma\,[\text{GeV}]$}
    \psfrag{anisotropy}[][]{\scriptsize $A$}
    \psfrag{flux}[bc][tc]{\scriptsize
    $dJ/dE_\gamma\,[\text{GeV}\text{cm}^{-2}\text{s}^{-1}\text{str}^{-1}]$}
    \psfrag{legendLong}[tc][cr]{
    \begin{minipage}[h]{2cm}\scriptsize
      DM$\rightarrow e^+e^-$
      \begin{align*}
        m_\text{dm}&=1000\GeV\\
        \tau_\text{dm}&=2.0\times10^{27}\s
      \end{align*}
    \end{minipage}}
    \psfrag{legendShort}[tc][cr]{\scriptsize DM$\rightarrow e^+e^-$}
    \psfrag{1090}[t][c]{\scriptsize$10^\circ:90^\circ$}
    \psfrag{1020}[t][c]{\scriptsize$10^\circ:20^\circ$}
    \psfrag{2060}[t][c]{\scriptsize$20^\circ:60^\circ$}
    \psfrag{6090}[t][c]{\scriptsize$60^\circ:90^\circ$}
    \includegraphics[width=0.88\linewidth]{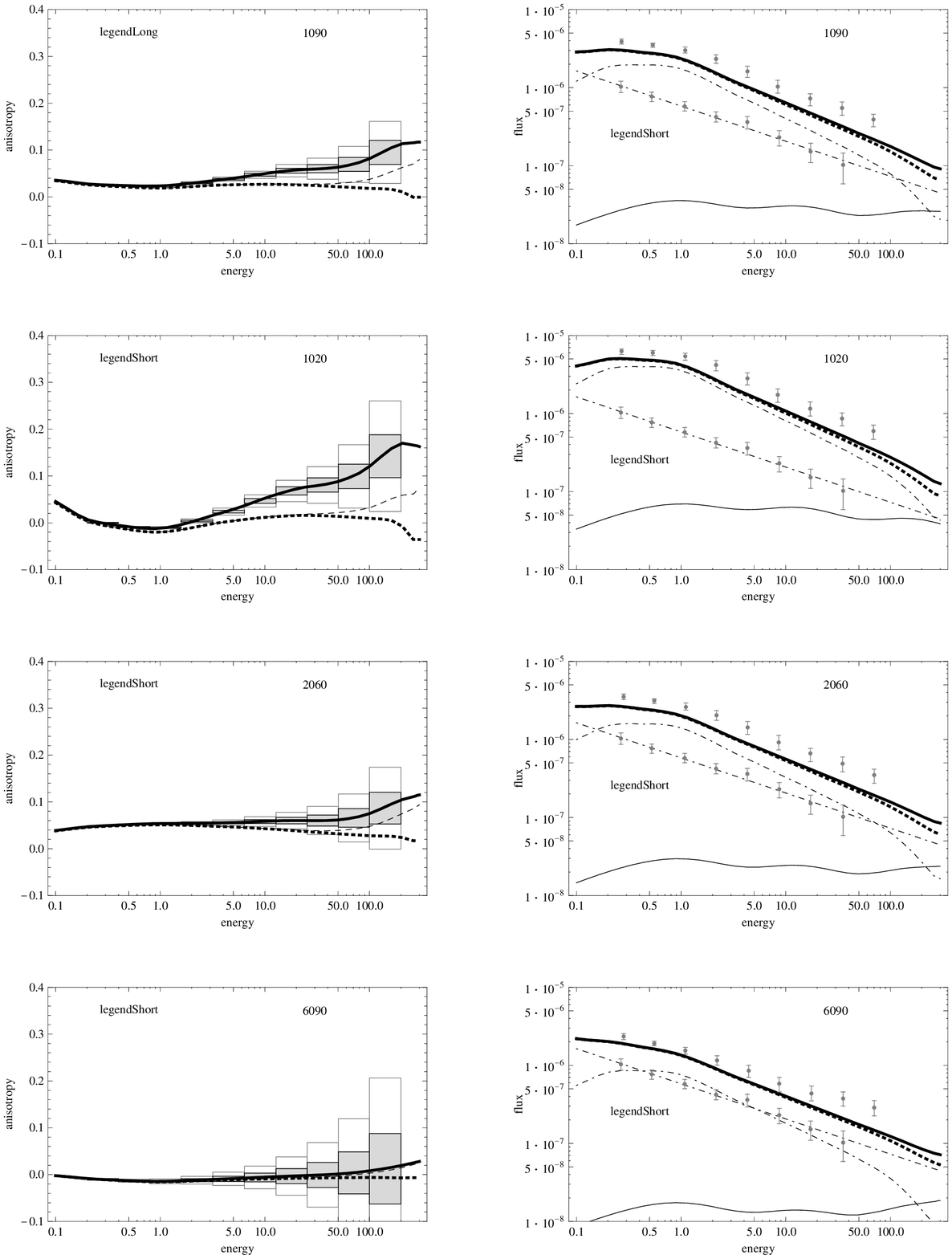}
  \end{center}
  \caption{Same as Fig.~\ref{fig:tautauAni}, but for decay into $e^+e^-$
  pairs. Here, in the \textit{left panels}, the difference in the predictions
  for the anisotropy when including ICS radiation (\textit{solid line}) and
  neglecting ICS radiation (\textit{dashed line}) is clearly visible.}
  \label{fig:eeAni}
\end{figure*}

To illustrate the impact of ICS radiation on the anisotropy parameter $A$, we
show in Fig.~\ref{fig:eeAni} the anisotropy of the gamma-ray flux assuming
that the dark matter particle decays into $e^+e^-$ pairs (with
$m_\text{dm}=1000\GeV$ and $\tau_\text{dm}=2\times10^{27}\s$). In this case
the dominant source of gamma rays is inverse Compton scattering (note that we
assumed that the decaying dark matter particle has spin 1; for scalar dark
matter particles, helicity suppression leads to an enhanced production of
final-state radiation \cite{Barger:2009xe}, weakening the relative
contribution from ICS). For reference, we also show the anisotropy that would
be measurable if ICS radiation were absent (dashed lines in left panels of
Fig.~\ref{fig:eeAni}).  Note that in this scenario electrons and positrons
produced in the dark matter decay give a sizeable contribution to the local
cosmic-ray fluxes, without being in conflict with the PAMELA and Fermi LAT
data. Again, we find that a sizable anisotropy is expected in several patches
of the sky. In this case, however, the gamma rays relevant for our predictions
are mainly produced close to the Galactic center, above and below the Galactic
disk. Hence, the anisotropies are relatively weak at higher latitudes
$|b|\gtrsim 20^\circ$. A more detailed discussion about the differences
between prompt and ICS gamma rays from dark matter decay can be found in
appx.~\ref{sec:Maps}.\\

\begin{figure*}[h!]
  \begin{center}
    \psfrag{energy}[][]{\scriptsize $E_\gamma\,[\text{GeV}]$}
    \psfrag{anisotropy}[][]{\scriptsize $A$}
    \psfrag{flux}[bc][tc]{\scriptsize
    $dJ/dE_\gamma\,[\text{GeV}\text{cm}^{-2}\text{s}^{-1}\text{str}^{-1}]$}

    \psfrag{legendLongTT}[tc][cr]{
    \begin{minipage}[h]{2cm}\scriptsize
      \hspace{-1cm}
      DM$\rightarrow \tau^+\tau^-$
      \begin{align*}
        m_\text{dm}&=5000\GeV\\
        \tau_\text{dm}&=1.2\times10^{26}\s
      \end{align*}
    \end{minipage}}
    \psfrag{legendShortTT}[tc][c]{\scriptsize DM$\rightarrow \tau^+\tau^-$}

    \psfrag{legendLongMM}[tc][cr]{
    \begin{minipage}[h]{2cm}\scriptsize
      \hspace{-1cm}
      DM$\rightarrow \mu^+\mu^-$
      \begin{align*}
        m_\text{dm}&=2500\GeV\\
        \tau_\text{dm}&=2.3\times10^{26}\s
      \end{align*}
    \end{minipage}}
    \psfrag{legendShortMM}[tc][c]{\scriptsize DM$\rightarrow \mu^+\mu^-$}

    \psfrag{legendLongWM}[tc][cr]{
    \begin{minipage}[h]{2cm}\scriptsize
      \hspace{-1cm}
      DM$\rightarrow W^\pm\mu^\mp$
      \begin{align*}
        m_\text{dm}&=3000\GeV\\
        \tau_\text{dm}&=2.7\times10^{26}\s
      \end{align*}
    \end{minipage}}
    \psfrag{legendShortWM}[tc][c]{\scriptsize DM$\rightarrow W^\pm\mu^\mp$}

    \psfrag{legendLongLLN}[tc][cr]{
    \begin{minipage}[h]{2cm}\scriptsize
      \hspace{-1cm}
      DM$\rightarrow \ell^+\ell^-\nu$
      \begin{align*}
        m_\text{dm}&=2500\GeV\\
        \tau_\text{dm}&=1.9\times10^{26}\s
      \end{align*}
    \end{minipage}}
    \psfrag{legendShortLLN}[tc][c]{\scriptsize DM$\rightarrow \ell^+\ell^-\nu$}

    \psfrag{legendLongMMN}[tc][cr]{
    \begin{minipage}[h]{2cm}\scriptsize
      \hspace{-1cm}
      DM$\rightarrow \mu^+\mu^-\nu$
      \begin{align*}
        m_\text{dm}&=3500\GeV\\
        \tau_\text{dm}&=1.4\times10^{26}\s
      \end{align*}
    \end{minipage}}
    \psfrag{legendShortMMN}[tc][c]{\scriptsize DM$\rightarrow \mu^+\mu^-\nu$}

    \psfrag{1090}[t][c]{\scriptsize$10^\circ:90^\circ$}
    \psfrag{1020}[t][c]{\scriptsize$10^\circ:20^\circ$}
    \psfrag{2060}[t][c]{\scriptsize$20^\circ:60^\circ$}
    \psfrag{6090}[t][c]{\scriptsize$60^\circ:90^\circ$}
    \includegraphics[width=0.88\linewidth]{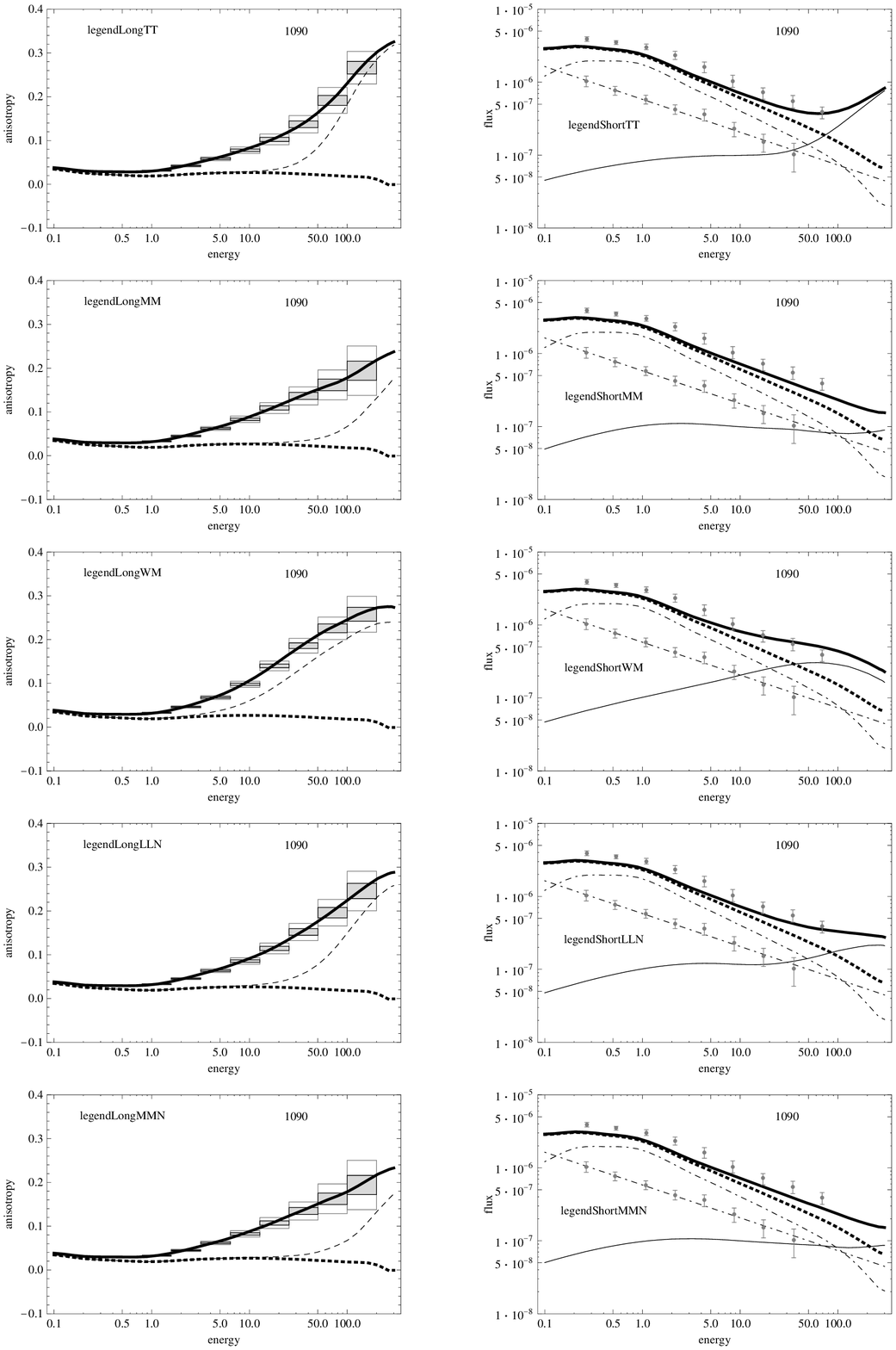}
  \end{center}
  \caption{Predictions for the five dark matter decay channels that where
  found to fit the positron excess as observed by PAMELA/Fermi. The
  \textit{left panels} show predictions for the total measurable anisotropy at
  latitudes $|b|>10^\circ$ (\textit{solid line}) and estimates of the
  corresponding 1-year and 5-year statistical error bars of Fermi LAT. The
  \textit{dotted} line is the anisotropy predicted by \textsc{Galprop}. In the
  \textit{right panels} we show predictions for the averaged fluxes in the
  same sky region. Line coding is the same as in Fig.~\ref{fig:tautauAni}.}
  \label{fig:pamAni}
\end{figure*}

If dark matter decay is the origin of the excess in the positron fraction
observed by PAMELA and in the total electron-plus-positron flux observed by
the Fermi LAT, the predicted anisotropies in the gamma-ray flux can be quite
large. In Fig.~\ref{fig:pamAni} we show our results for the anisotropy, $A$,
which is expected to be observed by the Fermi LAT if the dark matter particle
decays via one of the five different decay channels that were found in
Ref.~\cite{Ibarra:2009dr} to fit well the positron and electron
data.\footnote{We use slightly larger lifetimes than the ones given in
Ref.~\cite{Ibarra:2009dr} to account for the larger value for the local dark
matter density used in this work.} To minimize the statistical errors, we
concentrate on the region defined by $b_0=10^\circ$ and $b_1=90^\circ$ (see
appx.~\ref{sec:Maps}). As apparent from the plots, the predictions for some
decay channels (namely the decay into $W^\pm\mu^\mp$) are already in conflict
with the preliminary results of the Fermi LAT collaboration for the EGBG,
whereas other decay channels are marginally consistent (\fex~the decay into
$\mu^+\mu^-$ pairs). However, even for those channels which are compatible
with the data, sizeable anisotropies, around $A\simeq0.2-0.3$, are predicted
at energies $E_\gamma\simeq100\GeV$. This is significantly different from the
anisotropy expected for the astrophysical foreground. As indicated by our
estimates of the statistical error bars for one-year Fermi LAT data taking,
this deviation should be clearly visible in the upcoming results for the
diffuse gamma-ray sky. On the other hand, its non-observation would set very
strong constraints on the decaying dark matter interpretation of the positron
excess observed by PAMELA/Fermi LAT.

Uncertainties in the determination of the above large-scale anisotropy come
from different sources. When neglecting ICS radiation, the prediction of an
anisotropy between $0.2$ and $0.3$ in the dark matter signal at latitudes
$|b|>10^\circ$ is relatively robust. The main sources of uncertainty are the
profile of the Milky Way dark matter halo and its normalization. As discussed
above, the dependence on the profile is rather weak (\cf~Tab.~\ref{tab:Ani}).
Only in case of a much lower value for the local dark matter density, say
$\rho_\odot=0.2\GeV\cm^{-3}$, and only for gamma rays with energies
$E'_\gamma\lesssim 10\GeV$, the anisotropy can become as small as
$A\simeq0.15$. On the other hand, the size and anisotropy of the ICS radiation
from electrons and positrons originating from dark matter decay 
is plagued by many uncertainties like
the exact height of the diffusion zone, the distribution of the ISRF and the
size of the Galactic magnetic field. In general the ICS radiation, and hence
the overall anisotropy of the observed flux, becomes stronger if the height of
the diffusive halo is increased, but a detailed study of these uncertainties
is beyond the scope of this paper. Note, however, that the large scale
anisotropies predicted for the decay channels shown in Fig.~\ref{fig:pamAni}
are sizeable even if ICS radiation is neglected (dashed lines in left
panels).\\

Finally we will discuss in a more quantitative way the main prospects for the
Fermi LAT to detect gamma rays from dark matter decay through the observations
of large-scale anisotropies. To this end we will neglect inverse Compton
radiation from the electrons and positrons produced in the decay of the dark
matter particle, and we will assume perfect subtraction of the Galactic
foreground. The remaining flux is then expected to be constituted by the
isotropic EGBG, which is possibly contaminated by anisotropic radiation from
dark matter decay. For definiteness, we will assume that the remaining flux
follows the power law in Eq.~\eqref{eqn:EGBG}.

Provided that a fraction $f_\text{s}$ of the considered gamma-ray photons in a
given energy range is due to decaying dark matter, the measured anisotropy $A$
is given by
\begin{equation}
  A=f_\text{s} A_\text{s} + (1-f_\text{s})A_\text{bg}\;.
  \label{eqn:SplitA}
\end{equation}
Here, $A_\text{s}$ and $A_\text{bg}$ denote the anisotropy of the dark matter
signal, which can be read off from Tab.~\ref{tab:Ani}, and the anisotropy of
the astrophysical background, which is $A_\text{bg}=0$ in our case,
respectively. A possible detection at the $3\sigma$-level requires that
\begin{equation}
  f_\text{s}>\frac{3\,\sigma_A}{A_\text{s}-A_\text{bg}}\;,
  \label{eqn:Adet}
\end{equation}
where $\sigma_A$ denotes the standard deviation of the anisotropy $A$. It
depends on the total number of measured photons, $N_\gamma$ and can be
approximated by $\sigma_A\simeq N_\gamma^{-1/2}$. The adopted approximation
for $\sigma_A$ is better than $10\%$ as long as $|A|\lesssim0.3$ and $\sigma_A
\lesssim0.2$. See appx.~\ref{sec:dA} for a short discussion. The photon
number is given by
\begin{equation}
  N_\gamma = \varepsilon\cdot \Omega_\text{sky} 
  \int_{E_0}^{E_1} dE_\gamma \frac{dJ}{dE_\gamma}\;,
  \label{eqn:defN}
\end{equation}
where $\varepsilon$ denotes the experimentally given exposure (see above) and
$\Omega_\text{sky}$ is the solid angle of the observed sky, which is given by
$\Omega_\text{sky}=0.83\cdot 4\pi$ if the Galactic disk with $|b|<10^\circ$ is
excluded. Following Eq.~\eqref{eqn:EGBG} the Fermi LAT will detect
$N_\gamma\simeq3.0\times10^4$ ($N_\gamma\simeq1.1\times10^3$) photons with
energies $E_\gamma\geq10\GeV$ ($E_\gamma\geq100\GeV$) after five years of data
taking. Taking for definiteness $A_\text{s}=0.3$, this allows in principle a
$3\sigma$-detection of a dark matter contamination down to 
$f_\text{s}\simeq 6\%$ ($f_\text{s}\simeq 30\%$). Note, however, that 
additional statistical noise and
systematic uncertainties from point source subtraction and the determination
of the Galactic foreground are neglected and can reduce the sensitivity to
$f_\text{s}$ by factors of order one.\\

\section{Angular Power Spectrum of gamma rays from decaying dark matter}
\label{sec:power}
We will now discuss the angular power spectrum $C_\ell$ of gamma rays which
stem from the decay of dark matter particles (for the annihilation case, see
\fex~\cite{AK06, AKN+07, SiegalGaskins:2008ge, SiegalGaskins:2009ux, Ando09,
FPB+09}). On large angular scales the angular power spectrum is expected to be
completely dominated by the dipole-like asymmetry of the halo flux, which is
-- as discussed above -- due to the offset between the Sun and the Galactic
center.  However, on smaller scales spatial fluctuations of the dark matter
density can become relevant. These fluctuations are related to subhalos of our
own Galactic halo, and to the large-scale structure of the dark matter
distribution in the nearby Universe. The abundance and distribution of
subhalos and their impact on indirect searches for dark matter is still a
debated question~\cite{Pieri:2007ir, PLB+09}. Recent $N$-body simulations
suggest that inside a radius of $r=100\kpc$ around the Galactic center, the
fraction of the halo mass that is bound in subhalos can be as small as a few
percent~\cite{Springel:2008cc}. Such a small value would lead to a strong
suppression of subhalo-related signals in case of decaying dark matter. In
this work we will concentrate on the more robust predictions related to the
large-scale structures of the dark matter distribution, and we will assume a
smooth Milky Way halo when calculating the angular power spectrum.
Furthermore, we neglect ICS radiation throughout this section.\\

The fluctuations of the dark matter gamma-ray flux in a certain energy band
can be expanded in spherical harmonics $Y_{\ell m}$ with coefficients $a_{\ell
m}$ according to
\begin{equation}
  \delta J(\vec{n}) \equiv J(\vec{n})-\langle
  J\rangle =\langle J\rangle \sum_{\ell m} a_{\ell m}
  Y_{\ell m}(\vec{n})\;.
  \label{eqn:APSexpansion}
\end{equation}
Here $\langle J\rangle$ denotes the mean gamma-ray flux, averaged over the
whole sky, and the coefficients $a_{\ell m}$ can be calculated from
\begin{equation}
  a_{\ell m}=\langle J \rangle^{-1}
  \int d\Omega_{\vec{n}}\, \delta J(\vec{n})\, Y_{\ell m}^\ast (\vec{n})\;.
  \label{eqn:alm}
\end{equation}

Our goal is to calculate the angular power spectrum $C_\ell\equiv\langle
|a_{\ell m}|^2 \rangle$, which can be split up into an extragalactic and a
halo part according to
\begin{equation}
  C_\ell = f_\text{eg}^2 C_\ell^\text{eg} + (1-f_\text{eg})^2
  C_\ell^\text{halo}\;.
  \label{eqn:ClSum}
\end{equation}
Here, $f_\text{eg}$ denotes the fraction of dark matter photons with
extragalactic origin. Cross-correlations between the halo and the
extragalactic component vanish in our case.\\

Following Eq.~\eqref{eqn:alm}, the angular power spectrum of the halo
component, $C_\ell^\text{halo}$, can be calculated by
\begin{equation}
  C_\ell^\text{halo} \propto
  \left|
  \int_0^\pi d\psi\,\sin\psi\, P_\ell^{0}(\cos\psi)\; J_\text{halo}(\psi)
  \right|^2\;,
  \label{eqn:ClHalo}
\end{equation}
where the $P_\ell^m$ denote the associated Legendre functions,
$J_\text{halo}(\psi)$ is the angular profile of the halo flux as plotted in
Fig.~\ref{fig:AngDist}, and the power spectrum has to be normalized such that
$C_{0}^\text{halo}=4\pi$.

For the calculation of the extragalactic component of the angular power
spectrum, $C_\ell^\text{eg}$, it is convenient to define the window function
\begin{equation}
  W(z)=
  \frac{\Omega_\text{dm}\rho_\text{c}}{4\pi m_\text{dm} \tau_\text{dm} }
  \int_{E_0}^{E_1} dE_\gamma \,
  \frac{dN_\gamma}{dE_\gamma}[(z+1)E_\gamma]\,
  e^{-\tau(E_\gamma,z)}\;,
  \label{eqn:windowF}
\end{equation}
for which holds that $\langle J_\text{eg} \rangle =\int_0^\infty dz\,
W(z)/H(z)$ (\cf~Eq.~\eqref{eqn:GammaFluxEG} above). The window function mainly
depends on the energy spectrum of gamma rays from the dark matter decay
$dN_\gamma/dE_\gamma$, and on the optical depth $\tau(E_\gamma,z)$. Following
the Limber approximation, the angular power spectrum of the extragalactic
component can be determined by (for details see \fex~Ref.~\cite{AKN+07})
\begin{equation}
  C_\ell^\text{eg} = \langle J_\text{eg} \rangle^{-2} 
  \int_0^\infty \frac{dr}{r^2}
  W(z)^2 P\left( k=\frac{\ell}{r},z \right)\;.
  \label{eqn:ClForm}
\end{equation}
Here, $P(k,z)$ denotes the power spectrum of the dark matter distribution as a
function of redshift $z$, and the comoving coordinate $r$ is related to $z$ 
by $dr=dz/H(z)$. The dark matter power spectrum at $z=0$ can be derived
from $N$-body simulations or analytically in the Halo model, and we will adopt
results from Ref.~\cite{CBH+08}. The corresponding redshift-dependence follows
from the relation
\begin{equation}
  P(k,z)=P(k)_{z=0}\cdot D(z)^2\;,
  \label{eqn:LGF}
\end{equation}
where $D(z)\propto H(z)\int_z^\infty dz' (1+z')H(z')^{-3}$ is the
linear-growth factor, normalized such that $D(0)=1$~\cite{Cooray:2002dia}. In
principle the above set of equations allow a calculation of the angular power
spectrum of gamma rays for arbitrary gamma-ray spectra.\\

\begin{figure}[h!]
  \begin{center}
    \psfrag{z}[][b]{\scriptsize $z$}
    \psfrag{r}[][t]{\scriptsize $r\,[\text{Gpc}]$}
    \psfrag{W}[][t]{\scriptsize $W(z)/H(z)$}
    \psfrag{i}[][]{\scriptsize $100\GeV$}
    \psfrag{j}[][]{\scriptsize $10\GeV$}
    \psfrag{k}[][]{\scriptsize $1\TeV$}
    \includegraphics[width=0.9\linewidth]{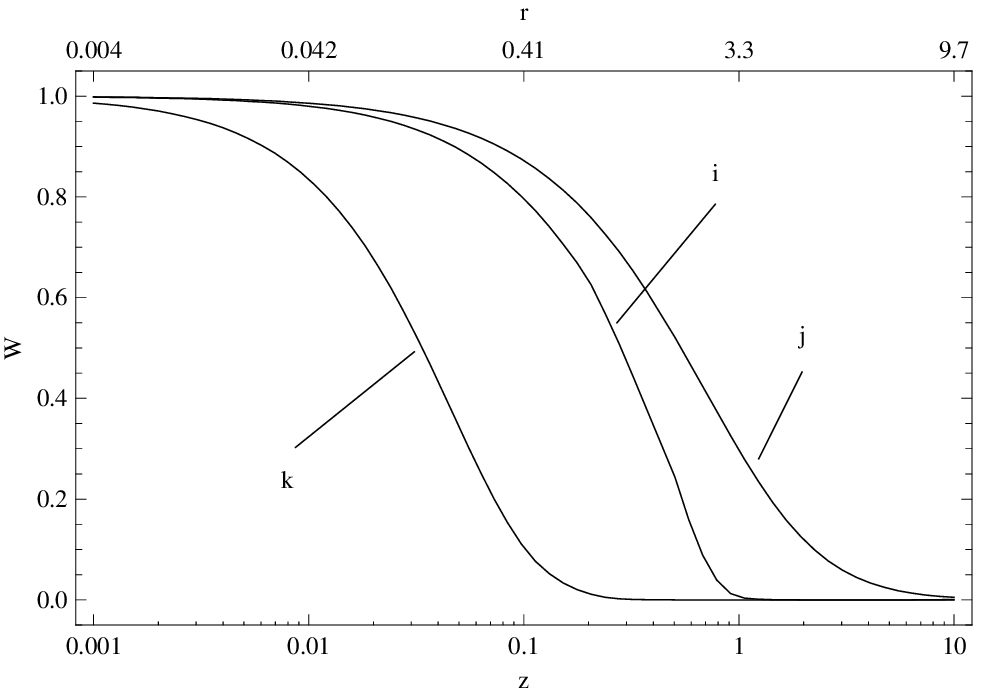}
  \end{center}
  \caption{Window function divided by Hubble rate, $W(z)/H(z)$, as a function
  of redshift $z$, with arbitrary normalization. We assume that dark matter
  decay is producing a monoenergetic line with energies between
  $E'_\gamma=10\GeV$ and $1\TeV$. For comparison, we also show values for the
  comoving coordinate $r(z)$.}
  \label{fig:Wfunc}
\end{figure}

In Fig.~\ref{fig:Wfunc} we show the quantity $W(z)/H(z)$ as a function of
redshift, assuming that dark matter decay produces a monoenergetic line with
energies between $E'_\gamma=10\GeV$ and $1\TeV$. When normalized
appropriately, $W(z)/H(z)\cdot dz$ gives the fraction of the received
extragalactic dark matter photons that were emitted at redshifts $z\dots
z+dz$. As evident from the figure, the region of the Universe that contributes
to the observable gamma-ray fluxes becomes larger for lower gamma-ray
energies. At $E'_\gamma=1\TeV$, around $90\%$ of the observable photons come
from comoving distances $r\lesssim400\Mpc$, whereas at lower gamma-ray
energies around $10\GeV$ photons are received that come from distances up to
$8\Gpc$. Hence, in the latter case fluctuations in the extragalactic dark
matter signal due to large-scale structures, which typically have sizes up to
several $100\Mpc$, are efficiently averaged out, which results in a
correspondingly small value of $C_\ell^\text{eg}$. For higher gamma-ray
energies the probed region shrinks, yielding a larger $C_\ell^\text{eg}$.
Note, however, that the fraction of gamma rays from the decay of dark matter
particles at cosmological distances, $f_\text{eg}$, presents an opposite
energy dependence and decreases by an order of magnitude, from $0.42$ to
$0.03$, when increasing the gamma-ray energy from $E'_\gamma=10\GeV$ to
$1\TeV$. This behavior counteracts the growth of $C_\ell^\text{eg}$ in
Eq.~\eqref{eqn:ClSum}.\\

\begin{figure}[h!]
  \begin{center}
    \psfrag{l}[][]{\scriptsize $\ell$}
    \psfrag{Cl}[][]{\scriptsize $\ell(\ell+1)C_\ell /2\pi$}
    \psfrag{p}[][]{\scriptsize photon noise}
    \psfrag{h}[][]{\scriptsize halo}
    \psfrag{e}[][]{\scriptsize extragalactic}
    \psfrag{i}[][]{\scriptsize $10\GeV$}
    \psfrag{j}[][]{\scriptsize $100\GeV$}
    \psfrag{k}[][]{\scriptsize $1\TeV$}
    \psfrag{b}[][]{\scriptsize $\geq100\GeV$}
    \psfrag{a}[][]{\scriptsize $\geq10\GeV$}
    \includegraphics[width=\linewidth]{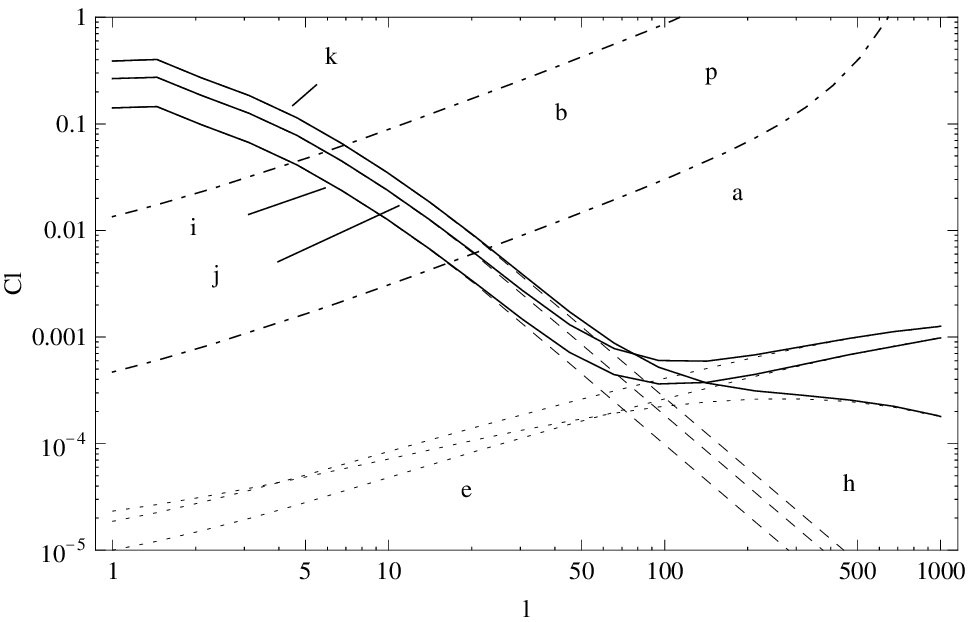}
  \end{center}
  \caption{Angular power spectrum of gamma-ray signal from decaying dark
  matter. Shown are the contributions from the halo component (\textit{dashed
  line}) and the contribution from the extragalactic radiation (\textit{dotted
  line}, based on results from \cite{CBH+08}). The \textit{dash-dotted} lines
  show for different gamma-ray energies our predictions of the photon noise
  level after five-years Fermi LAT observation.}
  \label{fig:AngPowSpec}
\end{figure}

Our results for the total angular power spectrum of gamma rays from decaying
dark matter are shown in Fig.~\ref{fig:AngPowSpec}, including the halo and the
extragalactic component, and assuming that the decay of the dark matter
particle only produces monoenergetic photons with energies between
$E'_\gamma=10\GeV$ and $1\TeV$. As apparent from the plot, the halo component
of $C_\ell$ increases slightly with $E'_\gamma$, which is related to a
corresponding decrease of $f_\text{eg}$ in Eq.~\eqref{eqn:ClSum}. The
variation of the extragalactic component with gamma-ray energy $E'_\gamma$ is
also on the $\mathcal{O}(1)$ level in most of the energy range, except at the
highest energies $E'_\gamma\gtrsim1\TeV$. In summary, the angular power
spectrum looks qualitatively similar in the whole considered energy regime.
The halo component dominates at large angles as long as $\ell\lesssim100$,
whereas the extragalactic component takes over at smaller angles with
$\ell\gtrsim100$, yielding values for the power spectrum around
$\ell(\ell+1)C_\ell/2\pi\simeq \mathcal{O}(10^{-4}-10^{-3})$.\\

To shortly discuss the observational prospects for Fermi LAT, we will simply
assume that the total EGBG (as given by Eq.~\eqref{eqn:EGBG}) above a certain
energy is due to dark matter decay. Then, following Ref.~\cite{AKN+07}, the
$1\sigma$ deviation of the power spectrum, averaged over $\ell$-bins with size
$\Delta\ell$, is given by
\begin{equation}
  \delta C_\ell = \sqrt{\frac{2}{(2\ell+1)\Delta \ell f_\text{sky}}}
  \left(C_\ell+\frac{C_N}{W_\ell^2} \right)\;,
  \label{eqn:APSnoise}
\end{equation}
where $C_N=4\pi f_\text{sky}N_\gamma^\text{tot}/(N_\gamma^\text{s})^2$ is the
photon noise, $N_\gamma^\text{s}$ and $N_\gamma^\text{tot}$ denote the number
of measured dark matter photons and number of total photons, respectively,
$f_\text{sky}$ is the observed fraction of the sky and
$W_\ell=\exp(-\sigma_b^2\ell^2/2)$ is the window function of a Gaussian
point-spread function. For the Fermi LAT we take as beam size
$\sigma_b=0.1^\circ$,\footnote{See \url{http://www-glast.slac.stanford.edu}}
and we set $\Delta\ell=0.5\,\ell$. We only consider latitudes with $|b|\geq
20^\circ$, which implies that $f_\text{sky}=0.66$. In this part of the sky the
ratio between the total gamma-ray fluxes and the EGBG flux is $\mathcal{O}(3)$
(see Fig.~\ref{fig:pamAni}), and we will set $N_\gamma^\text{tot}/N_\gamma^\text{s}=3$
for definiteness. The number of signal photons follows from
Eqs.~\eqref{eqn:GammaFluxEG}~and~\eqref{eqn:defN}. 

The resulting photon noise (second term of $\delta C_\ell$ in
Eq.~\eqref{eqn:APSnoise}) for five-years Fermi LAT observation is shown in
Fig.~\ref{fig:AngPowSpec} by the two dash-dotted lines, taking into account
only gamma rays with energies $E_\gamma\geq10\GeV$ or $\geq100\GeV$,
respectively. As apparent from these lines, the small-scale fluctuations are
completely hidden under photon noise in the energy range we consider, and
hence they are not accessible by current instruments like Fermi LAT. However,
note that this is in contrast to gamma-ray signals from annihilating dark
matter~\cite{SiegalGaskins:2009ux, FPB+09} or different extragalactic
astrophysical sources~\cite{Ando:2006mt, Miniati:2007ke}, where the
small-scale anisotropies can be orders of magnitude larger. This fact might
eventually provide a handle on distinguishing the different scenarios, but a
detailed investigation of the prospects is beyond the scope of this paper.

\section{Conclusions}
\label{sec:Conclusion}
Dark matter particles could decay into gamma rays at a rate which is
sufficiently large as to allow their indirect detection through an excess over
the expected power law in the energy spectrum of the diffuse extragalactic
gamma-ray background. In this work we have discussed a complementary way of
indirectly detecting unstable dark matter particles by exploiting the fact
that the Earth is not located at the center of the dark matter halo. We have
discussed the relative size of the extragalactic and the halo component of the
gamma rays from dark matter decay, incorporating the attenuation effects from
pair-production on the intergalactic background light, and we have calculated
the dipole-like anisotropy between the high latitude gamma-ray flux coming
from Galactic center and from Galactic anticenter regions for different decay
channels. We have furthermore demonstrated the strong impact of gamma-rays
from inverse Compton scattering of electrons and positrons from dark matter
decay on the anisotropy signal. We have found that if dark matter decay is the
correct explanation of the excesses in the positron fraction and the total
electron-plus-positron flux reported by the PAMELA and Fermi LAT
collaborations, such an anisotropy in the gamma-ray flux should be observed by
the Fermi LAT (see Fig.~\ref{fig:pamAni}). 

Lastly, we have calculated the angular power spectrum of the gamma-ray signal
from dark matter decay, which exhibits imprints from the large scale structure
of the Universe (see Fig.~\ref{fig:AngPowSpec}). We have shown, however, that
for gamma ray energies above a few GeV these small scale effects are too weak
to be observed with present instruments.

\begin{acknowledgments}
  C.W.~would like to thank Luca Maccione, G\"{u}nter Sigl and Le Zhang for
  helpful discussions. A.I. would like to thank the Korea Institute for
  Advanced Study (KIAS) for hospitality during the last stages of this work.
  The work of A.I. and D.T. was partially supported by the DFG cluster of
  excellence ``Origin and Structure of the Universe.'' 
\end{acknowledgments}

\appendix 
\section{Statistical Errors of the Large-Scale Anisotropy}
\label{sec:dA}
Here, we shortly discuss the statistical errors of the large-scale anisotropy
as defined in Eq.~\eqref{eqn:Ani}, which are due to shot noise. Statistical
errors of small-scale anisotropies are discussed \fex~in
Ref.~\cite{Cuoco:2006tr}.

The measured anisotropy $A$ and the total number of measured photons $N$ are
related to the number of photons measured in direction of the Galactic center,
$N_1$, and anticenter, $N_2$, by $A=(N_1-N_2)/(N_1+N_2)$ and $N=N_1+N_2$. The
$N_i$ follow a Poisson distribution with mean $\langle N_i\rangle$ and
standard deviation $\sigma_{N_i}=\sqrt{\langle N_i\rangle}$. Considering the
propagation of uncertainty, it is straightforward to derive that the
statistical error of the anisotropy is given by
\begin{equation}
  \sigma_A \simeq \sqrt{\frac{1-\langle A\rangle^2}{\langle N\rangle}}\;,
  \label{eqn:DeltaA}
\end{equation}
which is expected to hold for small enough $\langle A\rangle\simeq A $ and
large enough $\langle N\rangle\simeq N$.

On the other hand, one can derive the exact probability distribution function
of the anisotropy $A$ by starting with the above Poisson distributions for the
$N_i$, performing an appropriate redefinition of the parameters and
integrating out the total number of measured photons. The result is a function
of the mean values $\langle A\rangle$ and $\langle N\rangle$ and can be
written in the compact form
\begin{equation}
  pdf(A)=
  \frac{\langle N\rangle}{2\langle N_1\rangle!\langle N_2\rangle!}
  \left( \frac{1+A}{2} \right)^{\langle N_1\rangle}
  \left( \frac{1-A}{2} \right)^{\langle N_2\rangle}\;.
  \label{eqn:pdf}
\end{equation}
From this equation one can check, for example, that a normal distribution with
mean $\langle A\rangle$ and standard deviation as in Eq.~\eqref{eqn:DeltaA}
gives correct error bars at the $5\%$ level as long as $\langle A\rangle <
0.6$ and $\sigma_A<0.2$. For small enough anisotropies $A$, however, the
standard deviation is just given by $\sigma_A=\sqrt{N^{-1}}$ with good
accuracy, and we use this approximation throughout the work.

\section{Observation of Prompt and ICS Radiation from Dark Matter Decay} 
\label{sec:Maps}
In this appendix we will briefly discuss the differences between observing ICS
radiation from positrons and electrons from dark matter decay and observing
the gamma rays that come directly from the decay itself (prompt radiation like
internal bremsstrahlung) by means of signal-to-noise and signal-to-background
ratios.

Signal-to-noise ratios quantify the significance of a signal against
statistical noise. The signal-to-noise ratio $S/N$ of a dark matter signal
with respect to the background is given by
\begin{equation}
  \frac{S}{N}=\frac{N_{\gamma,\text{s}}}{\sqrt{N_{\gamma,\text{s}}+
  N_{\gamma,\text{bg}}}}\;, \label{eqn:SNdef}
\end{equation}
where $N_{\gamma,\text{s}}$ and $N_{\gamma,\text{bg}}$ denote the number of
detected signal and background photons, respectively, that are observed in a
given sky region $\Delta\Omega$ and energy band $E_\gamma=E_0\dots E_1$.
Signal photons are all photons from dark matter decay, background photons are
in principle all other observed photons, including the astrophysical part of
the EGBG and the Galactic foreground. Since the number of detected photons
scales like
$N_{\gamma,i}\propto\int_{\Delta\Omega}d\Omega\int_{E_0}^{E_1}dE\,dJ_i/dE_\gamma
\equiv\Delta\Omega\, \bar{J}_i$, the signal-to-noise ratio in the limit
$\Delta\Omega\rightarrow0$ is proportional to
\begin{equation}
  \frac{S}{N}\propto
  \frac{\bar{J}_\text{s}}{\sqrt{\bar{J}_\text{s}+\bar{J}_\text{bg}}}\;,
  \label{eqn:SNdefp}
\end{equation}
where $\bar{J}_\text{s}$ and $\bar{J}_\text{bg}$ denote the appropriately
averaged and integrated signal and background gamma-ray fluxes, respectively.

\begin{figure}[h!]
  \vspace{0.2cm}
  \begin{center}
    \psfrag{l}[][]{\scriptsize$l\,[\text{degree}]$}
    \psfrag{b}[][]{\scriptsize$b\,[\text{degree}]$}
    \psfrag{def}[][]{\scriptsize signal-to-noise ratio of prompt radiation}
    \psfrag{l4}[][]{\scriptsize $1$}
    \psfrag{l3}[][]{\scriptsize $0.37$}
    \psfrag{l2}[][]{\scriptsize $0.14$}
    \psfrag{l1}[][]{\scriptsize $0.05$}
    \includegraphics[width=\linewidth]{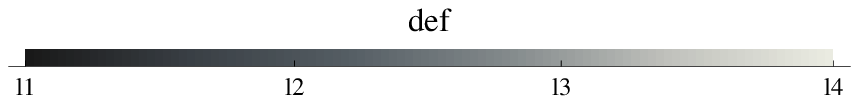}
    \includegraphics[width=\linewidth]{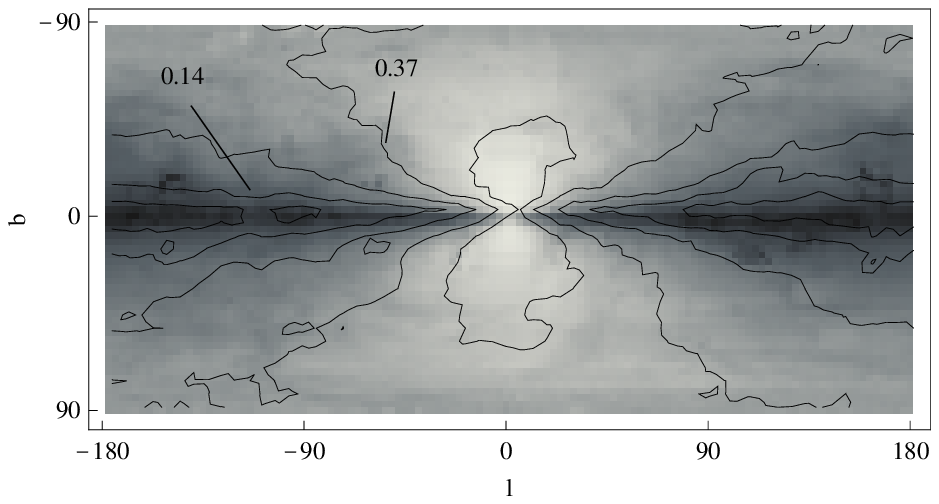}
    \psfrag{def}[][]{\scriptsize signal-to-background ratio of prompt
    radiation}
    \psfrag{l4}[][]{\scriptsize $1$}
    \psfrag{l3}[][]{\scriptsize $0.22$}
    \psfrag{l2}[][]{\scriptsize $0.05$}
    \psfrag{l1}[][]{\scriptsize $0.01$}
    \includegraphics[width=\linewidth]{pics/bar.eps}
    \includegraphics[width=\linewidth]{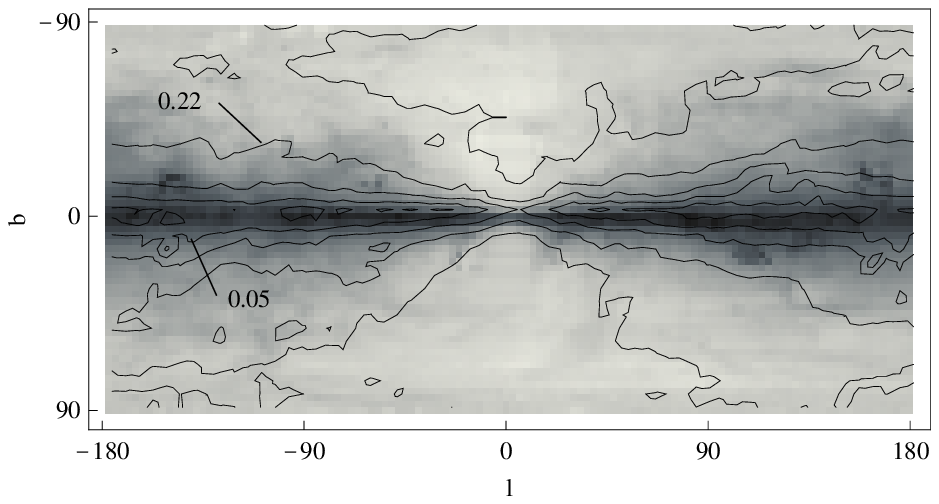}
  \end{center}
  \caption{\textit{Upper panel:} Relative signal-to-noise ratio of gamma-ray
  signal from dark matter decay as a function of the Galactic longitude $l$
  and latitude $b$, normalized to one at the Galactic center. \textit{Lower
  panel:} Relative signal-to-background ratio for the same process.
  Extragalactic and ICS radiation is neglected, and as background we take the
  predictions of the conventional \textsc{Galprop} model at $100\GeV$.}
  \label{fig:PRshape}
\end{figure}

In the upper panel of Fig.~\ref{fig:PRshape} we plot the relative
signal-to-noise ratio (with arbitrary normalization), assuming that the
background completely dominates the signal, $\bar{J}_\text{s}\ll
\bar{J}_\text{bg}$, and neglecting ICS and extragalactic radiation. As
background we take the predictions of the conventional \textsc{Galprop} model
at $E_\gamma=100\GeV$ (from Ref.~\cite{SMR04a}, see above), but the results do
not change qualitatively for other energies. The anisotropy of the dark matter
signal as function of $l$ is well recognizable in the plot. Furthermore it is
apparent that, from the perspective of statistical noise, the sky regions that
are most sensitive to decaying dark matter signals lie close above and below
the Galactic center, with $|l|\lesssim 25^\circ$ and
$5^\circ\lesssim|b|\lesssim 35^\circ$.

To determine the best observational strategy in light of the systematics that
are related to the determination of the Galactic foreground, it is more
convenient to consider the signal-to-background ratio $J_\text{s}/J_\text{bg}$
(we assume that systematic uncertainties scale roughly like $\sim
J_\text{bg}$). We show the signal-to-background ratio as a function of the
Galactic coordinates in the lower panel of Fig.~\ref{fig:PRshape}. Again, the
large-scale anisotropy of the dark matter flux are clearly visible in the
plot. Furthermore, it is apparent that concerning systematics the best
strategy is to avoid regions near the Galactic plane and to observe fluxes
only at higher latitudes, $|b|\gtrsim 20^\circ$. However, as a compromise
between statistical and systematic uncertainties we choose to consider the
whole region $10^\circ\leq|b|\leq90^\circ$ in most of the present paper.

\begin{figure}[h!]
  \vspace{0.2cm}
  \begin{center}
    \psfrag{l}[][]{\scriptsize$l\,[\text{degree}]$}
    \psfrag{b}[][]{\scriptsize$b\,[\text{degree}]$}
    \psfrag{def}[][]{\scriptsize signal-to-background ratio of ICS radiation}
    \psfrag{l4}[][]{\scriptsize $1$}
    \psfrag{l3}[][]{\scriptsize $0.20$}
    \psfrag{l2}[][]{\scriptsize $0.04$}
    \psfrag{l1}[][]{\scriptsize $0.008$}
    \includegraphics[width=\linewidth]{pics/bar.eps}
    \includegraphics[width=\linewidth]{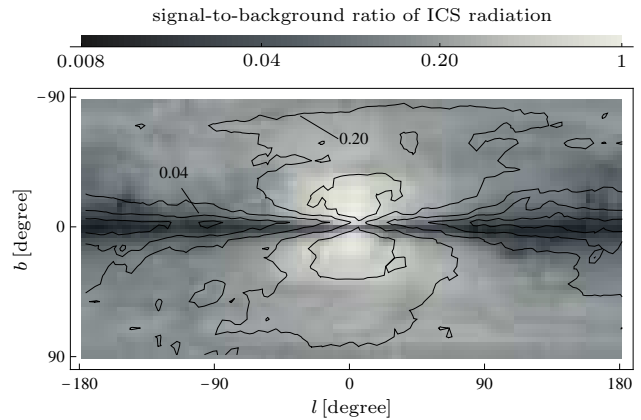}
  \end{center}
  \caption{Relative signal-to-background ratio of pure ICS radiation from dark
  matter decay at $E_\gamma=10\GeV$. We assume that $m_\text{dm}=1\TeV$ and
  that the dark matter is decaying into $e^+e^-$ pairs. As background we take
  the fluxes from the conventional \textsc{Galprop} model.}
  \label{fig:ICSshape}
\end{figure}

In contrast to gamma rays that come directly from the dark matter decay, the
gamma rays that stem from inverse Compton scattering of dark matter positrons
or electrons on the ISRF are mostly coming from the region near the Galactic
center. This can be seen in Fig.~\ref{fig:ICSshape}, where we plot the
signal-to-background ratio of the pure ICS signal of dark matter decaying into
an $e^+e^-$ pair (with $m_\text{dm}=1\TeV$) as a function of the Galactic
coordinates. The gamma-ray energy is $E_\gamma=10\GeV$ in this plot. As
background we again use the predictions from the conventional \textsc{Galprop}
model. From the figure it is apparent that the relative size of the signal
peaks at regions very close below and above the Galactic center, with
$|l|\lesssim20^\circ$ and $5^\circ\lesssim|b|\lesssim30^\circ$. This suggests
that concentrating the observation on these regions is most promising for the
search for ICS radiation from dark matter decay. However, in light of the
large underlying uncertainties related to the predictions of ICS radiation we
will neglect these subtleties, and we consider ICS radiation only in how far
it affects the anisotropies and fluxes in the sky regions that are most
promising for the search of gamma rays coming from the dark matter decay
itself.

\bibliography{}
\end{document}